\documentclass[pra,showpacs,amsmath,amssymb,twocolumn]{revtex4}
\usepackage{amssymb}
\usepackage{mathrsfs}
\usepackage{amsfonts}

\usepackage{amsthm}
\usepackage{dcolumn}
\usepackage{bm}
\usepackage{graphicx}

\newcommand\ket[1]{\ensuremath{|#1\rangle}}
\newcommand\bra[1]{\ensuremath{\langle#1|}}
\newcommand\iprod[2]{\ensuremath{\langle#1|#2\rangle}}

\begin{document}

\title{A Unified Quantum NOT Gate}

\author{Zong-Wen Yu$^1$, Xiao-Tong Ni$^1$, Leong Chuan Kwek$^2$ and Xiang-Bin
Wang$^{3,4}$}

\affiliation{$^1$ Department of Mathematical Sciences, Tsinghua
University, Beijing 100084, China\\ $^2$ Centre for Quantum
Technologies and Department of Physics, National University of
Singapore, 3 Science Drive 2, Singapore 117543, Singapore
\\ $^3$ Department of Physics, Tsinghua University, Beijing 100084,
China and
\\ $^4$ Tsinghua National Laboratory for Information Science and Technology, Beijing 100084,
China}


\begin{abstract}
We study the feasibility of implementing a quantum NOT gate
(approximate) when the quantum state lies between two latitudes on
the Bloch's sphere and present an analytical formula for the
optimized $1$-to-$M$ quantum NOT gate. Our result generalizes
previous results concerning quantum NOT gate for a quantum state
distributed uniformly on the whole Bloch sphere as well as the phase
covariant quantum state. We have also shown that such $1$-to-$M$
optimized NOT gate can be implemented using a sequential generation
scheme via matrix product states (MPS).
\end{abstract}

\pacs{03.67.Lx, 03.65.Bg, 03.65.-a}

\maketitle

\section{Introduction}\label{sec:secIntro}
Recent developments in quantum information have resulted in an
increasing number of applications: for instance, quantum
teleportation, quantum dense coding, quantum cryptography, quantum
logic gates, quantum algorithms and
etc~\cite{Nielsen2000,Galindo2002,Gisin and Thew,Gisin and
Ribordy,rev}. Many tasks in quantum information processing (QIP)
possess different properties from their classical
counterparts. One such case is quantum NOT gate.
Classically, we can use the NOT gate to invert (complement) a bit, by
changing the value of a bit, from 0 to 1 and vice versa.
Complementing a qubit, however, is another matter. The complement of
a state $\ket{\psi}$ is the state $\ket{\psi^{\bot}}$ that is
orthogonal to it. In the quantum case, as shown by Bu\v{z}ek,
Hillery and Werner~\cite{BHW-UNG1999}, it is impossible to build a
device that  transforms an unknown quantum state into the state
orthogonal to it. That is to say, we cannot design a perfect
universal-NOT (U-NOT) gate. This difference between
classical and quantum information processing is closely
related to the no-cloning theorem~\cite{Wootters1982}. However, such
no-go theorem does not forbid imperfect quantum quantum
cloning~\cite{Buzek1996, Gisin1997, Brub1998, Gisin1998, Werner1998,
Keyl1999, Brub2000, phase1, phase2, buc, phase3, phase4, phase5,
Zou2005, Cerf2002, Fan2003, Fan2001, Cerf and Durt2002, Durt2003,
Hu2008}. Also, approximate quantum NOT gates do
exist~\cite{BHW-UNG1999} and it is interesting to know how well we
can orthogonalize an unknown quantum state. Bu\v{z}ek, Hillery and
Werner~\cite{BHW-UNG1999} have introduced a U-NOT gate that implements an
approximate NOT operation to an unknown quantum state $\ket{\psi}$
on the Bloch's sphere and generates an output that is as close as
possible to $\ket{\psi^{\bot}}$, which is orthogonal to
$\ket{\psi}$.

In many real applications of the quantum information system, we
often have partial information about a  2-level quantum state,
i.e., the state is distributed on a specific area on the Bloch
sphere. Such partial information as in phase covariant
$1$-to-$1$ NOT gate allows us to orthogonalize such states by
transforming $\ket{0}$ to $\ket{1}$ and $\ket{1}$ to $-\ket{0}$.
Thus, any phase covariant states can be orthogonalized
perfectly. In this work, we study the problem of 1-to-$M$ quantum NOT gate where
the input state is uniformly distributed between two latitudes of
the Bloch sphere rather than the whole Bloch sphere. By considering the case in which the
two latitudes are brought to the poles, we obtain the U-NOT gate with the
optimal fidelity $F=2/3$~\cite{BHW-UNG1999}. However, if the two latitudes
collapse into a geodesic circle of the Bloch sphere, we
obtain the phase covariant NOT gate.

Taking qubit $\ket{\psi}=\cos\frac{\theta}{2}\ket{0}+
\sin\frac{\theta}{2} e^{i\phi}\ket{1}$ with $\phi\in[0,2\pi]$ and
$\theta_{1}\leq\theta\leq\theta_2$ as input, and the outputs of our
$1$-to-$M$ NOT gate will always be some multipartite entangled
states. As a result, the controllable generation of these entangled
states becomes very important. But in general, it is extremely
difficult to generate experimentally multipartite entangled states
through a single global unitary operation. For this purpose, the
sequential generation of the entangled states appears to be more promising
and a lot of effort has been made in recent years in this direction.
The general
sequential generation of entangled multiqubit states in the realm of
cavity QED have been systematically studied
in~\cite{SchonPRL2005,SchonPRA2007}. It is pointed out that the
$1$-to-$M$ sequential quantum cloning is
possible~\cite{DelgadoPRL2007}. Dang and Fan~\cite{DangJPA2008}
extended the sequential quantum cloning to the general $N$-to-$M$ case
and considered also $d$-level systems.

To this end, we consider the following state:
\begin{equation}
  \ket{\psi}=\cos{\frac{\theta}{2}}\ket{0}+
  \sin{\frac{\theta}{2}}e^{i\phi}\ket{1}
\end{equation}
where $\phi\in[0,2\pi]$ and $\theta_{1}\leq\theta\leq\theta_2$ with
constants $\theta_1,\theta_2$. The states we considered here are
distributed uniformly between two latitudes on the Bloch sphere.
When $\theta_{1}=0$ and $\theta_{2}=\pi$, we get the situation of
the U-NOT gate. In this way, the result of U-NOT gate is recovered as
special cases of our NOT gate. When $\theta_{1}=\pi/2$ and
$\theta_{2}=\pi/2$, we obtain the NOT gate for phase covariant
states.

This paper is arranged as follows: We formulate our problem and
present analytical results to our situation in the next section. In
Sec.\uppercase\expandafter{\romannumeral3}, we analyze the
$1$-to-$M$ NOT gate within a sequential generation scheme and
express the sequential NOT gate in explicit form. We end the paper
with some concluding remarks.

\section{Quantum NOT gate for qubits between two latitudes on the Bloch
sphere  }\label{sec:NG} The state we wish to orthogonalize can be
written as
\begin{equation}\label{eq:State}
  \ket{\psi}=\cos{\frac{\theta}{2}}\ket{0}+
  \sin{\frac{\theta}{2}}e^{i\phi}\ket{1}
\end{equation}
where $\phi\in[0,2\pi]$ and $\theta_1\leq\theta\leq\theta_2$, i.e.,
the states we considered here are distributed uniformly within a region enclosed by
two latitudes on the Bloch sphere. We assume
the following unitary transformation for our NOT gate:
\begin{eqnarray}\label{eq:eqNG}
  U: &&
  \ket{0}\ket{X}\rightarrow
  \sum_{k=0}^{M}{\ket{(M-k)0,k1}\otimes\ket{A_{k}}} \nonumber \\
  &&
  \ket{1}\ket{X}\rightarrow
  \sum_{k=0}^{M}{\ket{k0,(M-k)1}\otimes\ket{A_{M+k+1}}}
\end{eqnarray}
where $\ket{(M-k)0,k1}$ denotes symmetric and normalized states with
$M-k$ qubits in $\ket{0}$ and $k$ qubits in $\ket{1}$. This ensures a
symmetric NOT gate and that all the
first $M$ qubits at the output of the NOT gate are the same.
$\ket{A_{l}}(l=0,1,\cdots,2M+1)$ are unormalized states. Let
$a_{k,l}=\iprod{A_l}{A_k}$ and denote $a_k=\iprod{A_k}{A_k}$ for
short where $a_{k,l}$ are the parameters that we want to determine.

After applying the unitary operation $U$, we can get the following
state with the input qubit $\ket{\psi}$ described by
Eq.~\eqref{eq:State}:
\begin{eqnarray}\label{eq:Out}
  \ket{\psi_{\textrm{out}}} &&
    =\cos\frac{\theta}{2}\sum_{k=0}^{M}{\ket{(M-k)0,k1}\otimes\ket{A_{k}}}
    \nonumber\\
    & &+\sin\frac{\theta}{2}e^{i\psi}\sum_{k=0}^{M}{\ket{k0,(M-k)1}\otimes\ket{A_{M+k+1}}}
\end{eqnarray}
By taking partial trace, we obtain the reduced density matrix
$\rho_{k}$ for the $k$-th output qubit, and all the reduced density
matrix are the same for $k=1,2,\cdots,M$. With the reduced density
matrix $\rho_{k}$, we can calculate the fidelity:
\begin{eqnarray}\label{eq:Fidelity}
  & &
    F=\bra{\psi^{\perp}}\rho_{k}\ket{\psi^{\perp}}\nonumber\\
  &=&
    \sin^{2}\frac{\theta}{2}\cos^{2}\frac{\theta}{2}\left[
    \sum_{k=0}^{M-1}{\frac{M-k}{M}(a_{k}+a_{M+k+1})}\right.
    \nonumber \\
  & &
    \left.+2\sum_{k=0}^{M-1}{\frac{\sqrt{(M-k)(K+1)}}{M}
    \sqrt{|a_{M+k+1,M-k-1}|^{2}}}\right] \nonumber\\
  & &
    +\sin^{4}\frac{\theta}{2}\sum_{k=0}^{M-1}
    {\frac{\binom{M-1}{M-k-1}}{\binom{M}{k+1}}a_{M+k+2}} \nonumber\\
  & &
    +\cos^{4}\frac{\theta}{2}\sum_{k=0}^{M-1}
    {\frac{\binom{M-1}{k}}{\binom{M}{k+1}}a_{k+1}} \nonumber\\
  & &
    +e^{i\phi}C_{1}+e^{-i\phi}C_{1}^{*}+e^{2i\phi}C_{2}+e^{-2i\phi}C_{2}^{*}
\end{eqnarray}
where $C_{1}^{*}$ is the complex conjugation of $C_{1}$ and the same
for $C_{2}^{*}$. sincee the input state $\ket{\psi}$ given by
Eq.~\eqref{eq:State} is arbitrary, the parameters $\phi\in [0,2\pi]$
and $\theta\in[ \theta_1, \theta_2]$ are unknown and are
distributed uniformly on a belt of the Bloch sphere.
We need to average the
fidelity over all possible cases. The last four terms in
Eq.~\eqref{eq:Fidelity} disappear as a consequence of the averaging
over all
possible angles $\phi$. Moreover, we obtain the same optimal NOT gate
by assuming that $C_{1}$ and $C_{2}$ are equal to zero. On the other
hand, by using the definitions of $a_{k,l}$ we can easily get that
$a_{k,l}=a_{l,k}^{*}$, $|a_{k,l}|^{2}= a_{k,l}*a_{l,k}\leq
a_{k}a_{l}$. Equality is obtained if and only if $a_{k,l}=a_{l,k}$ are
real numbers.  Using Eq.~\eqref{eq:Fidelity}, the fidelity becomes:
\begin{eqnarray}\label{eq:SimpleF}
    F
  &=&
    \sin^{2}\frac{\theta}{2}\cos^{2}\frac{\theta}{2}\left[
    \sum_{k=0}^{M-1}{\frac{M-k}{M}(a_{k}+a_{M+k+1})}\right.
    \nonumber \\
  & &
    \left.+2\sum_{k=0}^{M-1}{\frac{\sqrt{(M-k)(K+1)}}{M}
    \sqrt{a_{M+k+1}a_{M-k-1}}}\right] \nonumber\\
  & &
    +\sin^{4}\frac{\theta}{2}\sum_{k=0}^{M-1}
    {\frac{\binom{M-1}{M-k-1}}{\binom{M}{k+1}}a_{M+k+2}} \nonumber\\
  & &
    +\cos^{4}\frac{\theta}{2}\sum_{k=0}^{M-1}
    {\frac{\binom{M-1}{k}}{\binom{M}{k+1}}a_{k+1}}
\end{eqnarray}
Averaging the fidelity over all possible angles
$\theta$~\cite{Gisin1997}, and observing that $\sum_{k=0}^{M}{a_k}=
\sum_{k=0}^{M}{a_{M+k+1}}=1$, we have
\begin{eqnarray}\label{eq:MF}
  \bar{F}
  &=&
    \frac{\int_{\theta_{1}}^{\theta_{2}}{F\sin{\theta}d{\theta}}}
    {\int_{\theta_{1}}^{\theta_{2}}{\sin{\theta}d{\theta}}}
    \nonumber\\
  &=&
    \frac{1}{2}+\frac{1}{6}K
    \nonumber\\
  & &
    +P\sum_{k=0}^{M-1}
    {\frac{\sqrt{(M-k)(k+1)}}{M}\sqrt{a_{M+k+1}a_{M-k-1}}} \nonumber\\
  & &
    -Q\sum_{k=0}^{M-1}{\frac{M-k}{M}a_{k}}
    -R\sum_{k=0}^{M-1}{\frac{M-k}{M}a_{M+k+1}}
\end{eqnarray}
where
$K=\cos^{2}{\theta_{1}}+\cos{\theta_1}\cos{\theta_2}+\cos^{2}{\theta_2}$,
$P=\frac{3-K}{6}$,
$Q=\frac{K}{6}+\frac{1}{4}\left(\cos{\theta_1}+\cos{\theta_2}\right)$,
$R=\frac{K}{6}-\frac{1}{4}\left(\cos{\theta_1}+\cos{\theta_2}\right)$,
and $K,P,Q,R$ are constants with given $\theta_1$ and $\theta_2$. In
order to get the optimal quantum NOT gate, we should maximize
$\bar{F}$ with respect to $a_k(k=0,1,\cdots,M-1,M+1,M+2,\cdots,2M)$.

We now seek a solution of $a_k$ with maximum $\bar{F}$. It is interesting to note that
if the state lies somehwere  on the whole Bloch sphere, $\theta_{1}=0$ and $\theta_2=\pi$, and we have
$K=1$, $P=\frac{1}{3}$ and $Q=R=\frac{1}{6}$. The optimal fidelity
is $\bar{F}=\frac{2}{3}$ with $a_{M+k+1}=a_{M-k-1}
(k=0,1,\cdots,M-1)$, recovering the well known result for the
$1$-to-$M$ U-NOT gate in~\cite{BHW-UNG1999}. In this case, the
fidelity is constant and the optimal Universal NOT gate can be
realized via a "measurement + re-preparation"
scheme~\cite{BHW-UNG1999}. In the general situation, with states
uniformly distributed in a belt on the Bloch sphere, the fidelity is
dependent on the number of output qubits $M$ as shown below in this
article. So we can not realized via a "measurement + re-preparation"
scheme for the general case.

For the case in which the state is phase covariant,
$\theta_1=\theta_2=\frac{\pi}{2}$, and we have
$K=0$, $P=\frac{1}{2}$ and $Q=R=0$. The optimal fidelity is
$\bar{F}=\frac{1}{2}+\frac{\sqrt{M(M+2)}}{4M}$ for even $M$, and
$\bar{F}=\frac{1}{2}+\frac{M+1}{4M}$ for odd $M$. This fidelity is
just equal to the fidelity of optimal $1$ to $M$ phase-covariant
quantum cloning machine~\cite{Brub2000,Fan2001,Scarani2005}. As
mentioned above, the $1$-to-$1$ phase-covariant NOT gate can be
constructed perfectly. So we can achieve the $1$-to-$M$ optimal
phase-covariant NOT gate by combining the $1$-to-$1$ perfect NOT
gate with the $1$-to-$M$ optimal phase-covariant cloning machine.
The fidelities of the $1$-to-$M$ optimal phase-covariant NOT
gate and QCM must be the same as analyzed before.

In the general situation, we need to optimize the fidelity in
Eq.~\eqref{eq:MF} under the restrictions $0\leq a_{k}\leq 1
(k=0,1,\cdots,M-1,M+1,\cdots,2M),$  $\sum_{k=0}^{M-1}{a_k}\leq 1,$
and $\sum_{k=0}^{M-1}{a_{M+k+1}}\leq 1.$ By considering the smoothness
of $\bar{F}$, the maximum value should be achieved at the extremal
points or on the boundary. We analyze the optimization problem with
restrictions and get the following optimal NOT gate for the following
situations:

\begin{enumerate}
  \item When $|\theta_{1}-\frac{\pi}{2}|\geq
  |\theta_{2}-\frac{\pi}{2}|$ and $M$ is odd, we have
  $a_{\frac{M-1}{2}}=\min((\frac{P}{2Q})^2,1),
  a_{M}=1-a_{\frac{M-1}{2}}, a_{\frac{3M+1}{2}}=1,
  a_{\frac{3M+1}{2},\frac{M-1}{2}}=a_{\frac{M-1}{2},\frac{3M+1}{2}}=
  -\sqrt{a_{\frac{M-1}{2}}}$, and $a_{k,l}=0$ otherwise. The
  fidelity is $\bar{F}=\frac{1}{2}+\frac{K}{6}+\frac{M+1}{2M}(
  \frac{P^{2}}{4Q}-R)$ for
  $a_{\frac{M-1}{2}}=(\frac{P}{2Q})^{2}$, and
  $\bar{F}=\frac{1}{2}+\frac{K}{6}+\frac{M+1}{2M}(P-Q-R)$ for
  $a_{\frac{M-1}{2}}=1$.
  \item When $|\theta_{1}-\frac{\pi}{2}|<
  |\theta_{2}-\frac{\pi}{2}|$ and $M$ is odd, we have
  $a_{\frac{M-1}{2}}=1,
  a_{\frac{3M+1}{2}}=\min((\frac{P}{2R})^2,1),
  a_{2M+1}=1-a_{\frac{3M+1}{2}},
  a_{\frac{3M+1}{2},\frac{M-1}{2}}=a_{\frac{M-1}{2},\frac{3M+1}{2}}=
  -\sqrt{a_{\frac{3M+1}{2}}}$, and $a_{k,l}=0$ otherwise. The
  fidelity is $\bar{F}=\frac{1}{2}+\frac{K}{6}+\frac{M+1}{2M}(
  \frac{P^{2}}{4R}-Q)$ for
  $a_{\frac{3M+1}{2}}=(\frac{P}{2R})^{2}$, and
  $\bar{F}=\frac{1}{2}+\frac{K}{6}+\frac{M+1}{2M}(P-Q-R)$ for
  $a_{\frac{3M+1}{2}}=1$.
  \item When $|\theta_{1}-\frac{\pi}{2}|\geq
  |\theta_{2}-\frac{\pi}{2}|$ and $M$ is even, we have
  $a_{\scriptstyle{\frac{M}{2}}}=\min(\frac{P^{2}(M+2)}{4Q^{2}M},1),
  a_{\scriptstyle{M}}=1-a_{\scriptstyle{\frac{M}{2}}},
  a_{\scriptstyle{\frac{3M}{2}}}=1,
  a_{\scriptstyle{\frac{3M}{2},\frac{M}{2}}}=
  a_{\scriptstyle{\frac{M}{2},\frac{3M}{2}}}=
  -\sqrt{a_{\scriptstyle{\frac{M}{2}}}}$, and $a_{k,l}=0$ otherwise.
  The fidelity is $\bar{F}=\frac{1}{2}+\frac{K}{6}+
  \frac{M+2}{2M}(\frac{P^{2}}{4Q}-R)$ for
  $a_{\scriptstyle{\frac{M}{2}}}=\frac{P^{2}(M+2)}{4Q^{2}M}$, and
  $\bar{F}=\frac{1}{2}+\frac{K}{6}+
  P\frac{\sqrt{\frac{M}{2}(1+\frac{M}{2})}}{M}-
  R\frac{M+2}{2M}-\frac{1}{2}Q$ for
  $a_{\scriptstyle{\frac{M}{2}}}=1$.
  \item When $|\theta_{1}-\frac{\pi}{2}|<
  |\theta_{2}-\frac{\pi}{2}|$ and $M$ is even, we have
  $a_{\frac{M}{2}-1}=1,
  a_{\frac{3M}{2}+1}=\min(\frac{P^{2}(M+2)}{4R^{2}M},1),
  a_{2M+1}=1-a_{\frac{3M}{2}+1},
  a_{\frac{M}{2}-1,\frac{3M}{2}+1}=
  a_{\frac{3M}{2}+1,\frac{M}{2}-1}= -\sqrt{a_{\frac{3M}{2}+1}}$, and
  $a_{k,l}=0$ otherwise. The fidelity is
  $\bar{F}=\frac{1}{2}+\frac{K}{6}+\frac{M+2}{2M}(\frac{P^{2}}{4R}-Q)$
  for $a_{\frac{3M}{2}+1}=\frac{P^{2}(M+2)}{4R^{2}M}$, and
  $\bar{F}=\frac{1}{2}+\frac{K}{6}+
  P\frac{\sqrt{\frac{M}{2}(1+\frac{M}{2})}}{M}-
  Q\frac{M+2}{2M}-\frac{1}{2}R$ for $a_{\frac{3M}{2}+1}=1$.
\end{enumerate}
The explicit NOT gate transformations have already been
presented in Eq.~\eqref{eq:eqNG} by letting
$\ket{A_{[\frac{M}{2}]}}=-\sqrt{a_{[\frac{M}{2}]}}\ket{\uparrow},
\ket{A_{M}}=\sqrt{1-a_{[\frac{M}{2}]}}\ket{\downarrow},
\ket{A_{[\frac{3M+1}{2}]}}=\ket{\uparrow}$, and $\ket{A_{k}}=0$
otherwise for case 1 and 3; by letting
$\ket{A_{[\frac{M-1}{2}]}}=\ket{\uparrow},
\ket{A_{[\frac{3M}{2}+1]}}=-\sqrt{a_{[\frac{3M}{2}+1]}}\ket{\uparrow},
\ket{A_{2M+1}}=\sqrt{1-a_{[\frac{3M}{2}+1]}}\ket{\downarrow}$, and
$\ket{A_{k}}=0$ otherwise for case 2 and 4.

It is interesting to note that the output states of the NOT gate given
by Eq.~\eqref{eq:eqNG} are always entangled. As shown by Bu\v{z}ek,
Hillery and Werner~\cite{BHW-UNG1999}, the optimal U-NOT gate can be
realized via a ``measurement + repreparation" scheme. Moreover for each
measurement result obtained, the prepared state can be taken to be a
product one. In this case, it is an ``easy" operations, and requires
no generation of entanglement, nor unitary operations - just
measurement and preparation of product states. Unfortunately, there
is no ``measurement + repreparation" scheme in the general situation,
as the optimal fidelity is dependent on the number of output qubits
$M$. So the generation of entanglement is unavoidable. As a result,
the controlled generation of these entangled states becomes very
important. In the next section, we consider the generation of these
entangled states and present the sequential quantum NOT gate.

\section{The $1$-to-$M$ Sequential Quantum NOT gate}\label{sec:SNG}

As shown in~\cite{SchonPRL2005,DelgadoPRL2007,DangJPA2008}, the
sequential generation of a multiqubit state is as follows. Let
$\mathcal{H}_{A}\simeq\mathbb{C}^{D}$ and $\mathcal{H}_{B}\simeq
\mathbb{C}^{2}$ be the Hilbert spaces characterizing a
$D$-dimensional ancillary system and a single qubit respectively. At
every step of the sequential generation of a multiqubit state, a
unitary time evolution will be acting on the joint system
$\mathcal{H}_{A}\otimes\mathcal{H}_{B}$. Assuming that each qubit is
initially in the state $\ket{0}$, we disregard the qubit at the
input and write the evolution in the form of an isometry
$V:\mathcal{H}_{A}\to \mathcal{H}_{A}\otimes\mathcal{H}_{B}$, where
$V=\sum_{i,\alpha,\beta}{V^{i}_{\alpha,\beta}\ket{\alpha,i}\bra{\beta}}$,
each $V^{i}$ is a $D\times D$ matrix and the isometry condition
takes the form $\sum_{i=0}^{1}{[V^{i}]^{\dagger}V^{i}}=1$. If we
apply successively $n$ operations of this form to an initial state
$\ket{\varphi_{I}}\in\mathcal{H}_{A}$, we obtain the state
$\ket{\Phi}=V^{[n]}\cdots V^{[2]}V^{[1]}\ket{\varphi_{I}}$. The $n$
generated qubits are in general entangled. Assuming in the last step
the ancilla decouples from the system, such that
$\ket{\Phi}=\ket{\varphi_{F}}\otimes\ket{\varphi}$, and we are left with
the $n$-qubit state
\begin{equation}\label{eq:eqMPS}
  \ket{\varphi}=\sum_{i_{1},\cdots,i_{n}=0}^{1}{\bra{\varphi_{F}}V^{[n]i_{n}}\cdots
  V^{[1]i_{1}}\ket{\varphi_{I}}\ket{i_{n}\cdots i_{1}}},
\end{equation}
where $\ket{\varphi_{F}}$ is the final state of the ancilla. The
state~\eqref{eq:eqMPS} is a matrix-product state (MPS) (cf., e.g.,
\cite{PerezQIC2007}, and references therein), already
comprehensively studied in~\cite{SchonPRL2005, SchonPRA2007,
DelgadoPRL2007, DangJPA2008, PerezQIC2007, AffleckPRL1987,
VidalPRL2003, VerstraetePRL2004}. Moreover, it was proven that any
multiqubit MPS can be sequentially generated~\cite{SchonPRL2005}.

The NOT gate given by Eq.~\eqref{eq:eqNG} can approximately
orthogonalize one input state to $M$ copies. Next we show that this
general $1$-to-$M$ NOT gate can be generated through a sequential
procedure. The basic idea is to show that the final states
$\ket{\Psi_{1M}^{k}}$ in Eq.~\eqref{eq:eqNG} can be expressed in its
MPS form. As presented in~\cite{SchonPRL2005}, any MPS can be
sequentially generated. We shall follow the method as
in~\cite{DelgadoPRL2007, DangJPA2008, VidalPRL2003}.

Taking one output entangled state in case 1 for example. We have
\begin{equation}\label{eq:OutExample}
  \ket{\Psi_{1M}^{0}}=-\sqrt{\gamma}
  \ket{\frac{M+1}{2}0,\frac{M-1}{2}1}\ket{0}+\sqrt{1-\gamma}\ket{M1}\ket{1}
\end{equation}
where $\gamma=a_{\frac{M-1}{2}}$. By Schmidt decomposition, we first
express the quantum state $\ket{\Psi_{1M}^{0}}$ as a bipartite pure
state in $\mathcal{H}_{A_{1}}\otimes\mathcal{H}_{B_{1}}$ with two
particle sets $A_{1}=\{1\}$ and $B_{1}=\{2,3,\cdots,M+1\}$.
\begin{eqnarray}\label{eq:eqSD1}
  \ket{\Psi_{1M}^{0}}
  &=&
    \lambda_{1}^{[1]}\ket{0}\ket{\psi_{1}^{[2,\cdots,M+1]}}+
    \lambda_{2}^{[1]}\ket{1}\ket{\psi_{2}^{[2,\cdots,M+1]}}\nonumber
    \\
  &=&
    \sum_{\alpha_{1},i_{1}}{\Gamma_{\alpha_{1}}^{[1]i_{1}}
    \lambda_{\alpha_{1}}^{[1]}\ket{i_{1}}\ket{\psi_{\alpha_{1}}^{[2,\cdots,M+1]}}},
\end{eqnarray}
where $\lambda_{\alpha_{1}}^{[1]}$ are eigenvalues of the first
qubit reduced density operator and we find
$\lambda_{1}^{[1]}=\sqrt{\gamma\frac{M+1}{2M}}$,
$\lambda_{2}^{[1]}=\sqrt{1-\gamma\frac{M+1}{2M}}$. Matching indices
in Eq~\eqref{eq:eqSD1}, we have
$\Gamma_{\alpha_{1}}^{[1]0}=\delta_{\alpha_{1},1}$ and
$\Gamma_{\alpha_{1}}^{[1]1}=\delta_{\alpha_{1},2}$. To correspond
with the MPS form in Eq.~\eqref{eq:eqMPS}, we define
\begin{equation}\label{eq:eqV1}
  V_{\alpha_{1}}^{[1]i_{1}}=\Gamma_{\alpha_{1}}^{[1]i_{1}}\lambda_{\alpha_{1}}^{[1]}.
\end{equation}

By successive Schmidt decomposition, the quantum state
$\ket{\Psi_{1M}^{0}}$ in Eq.~\eqref{eq:OutExample} can be considered
as a bipartite pure state in
$\mathcal{H}_{A_{n}}\otimes\mathcal{H}_{B_{n}}$ with particle sets
$A_{n}=\{1,2,\cdots,n\}$ and $B_{n}=\{n+1,n+2,\cdots,M+1\}$, where
$1<n\leq M$. We have
\begin{equation}\label{eq:eqSDn}
  \ket{\Psi_{1M}^{0}}= \sum_{l=0}^{n}{ \lambda_{l+1}^{[n]}
  \ket{(n-l)0,l1}\ket{\psi_{l+1}^{[n+1,\cdots,M+1]}}},
\end{equation}
where $\lambda_{l+1}^{[n]}$ are eigenvalues of the first $n$ qubits
reduced density operator of $\ket{\Psi_{1M}^{0}}$. We can obtain
\begin{widetext}
\begin{equation}\label{eq:eqLambdan}
  \left\{
  \begin{array}{ll}
    \lambda_{l+1}^{[n]}=\sqrt{\gamma\frac{\binom{n}{l}
    \binom{M-n}{\frac{M-1}{2}-l}}{\binom{M}{\frac{M-1}{2}}}}, &
    1<n\leq\frac{M-1}{2}, l=0,1,\cdots,n-1 ;\quad
    \frac{M+1}{2}\leq n\leq M,
    l=n-\frac{M+1}{2},\cdots,\frac{M-1}{2}.\\
    \lambda_{n+1}^{[n]}=\sqrt{1-\gamma+\gamma
    \frac{\binom{M-n}{\frac{M-1}{2}-n}}{\binom{M}{\frac{M-1}{2}}}},
    & \begin{array}{lrl}
      1<n\leq \frac{M-1}{2}. &
      \hspace{5mm}\lambda_{n+1}^{[n]}=\sqrt{1-\gamma}, &
      \frac{M+1}{2}\leq n\leq M.
    \end{array} \\
    \lambda_{l+1}^{[n]}=0, & \textrm{otherwise}.
  \end{array}
  \right.
\end{equation}
and
\begin{equation}\label{eq:eqReducedState}
  \left\{
  \begin{array}{ll}
    \ket{\psi_{l+1}^{[n+1,\cdots,M+1]}}=
    -\ket{(\frac{M+1}{2}-n+l)0,(\frac{M-1}{2}-l)1}\ket{0}, &
    \hspace{-40mm} 1<
    n\leq \frac{M-1}{2},l=0,1,\cdots,n-1. \\
    \ket{\psi_{n+1}^{[n+1,\cdots,M+1]}}=
    -\frac{\sqrt{\gamma}}{\lambda_{n+1}^{[n]}}
    \sqrt{\frac{\binom{M-n}{\frac{M-1}{2}-n}}{\binom{M}{\frac{M-1}{2}}}}
    \ket{\frac{M+1}{2}0,(\frac{M-1}{2}-n)1}\ket{0}+
    \frac{\sqrt{1-\gamma}}{\lambda_{n+1}^{[n]}}\ket{(M-n)1}\ket{1},
    & 1<n\leq \frac{M-1}{2}. \\
    \ket{\psi_{l+1}^{[n+1,\cdots,M+1]}}=
    -\ket{(\frac{M+1}{2}-n+l)0,(\frac{M-1}{2}-l)1}\ket{0}, &
    \hspace{-40mm} \frac{M+1}{2}\leq n\leq M,
    l=n-\frac{M+1}{2},\cdots,\frac{M-1}{2}. \\
    \ket{\psi_{n+1}^{[n+1,\cdots,M+1]}}= \ket{(M-n)1}\ket{1}, &
    \hspace{-40mm} \frac{M+1}{2}\leq n\leq M.\\
    \ket{\psi_{l+1}^{[n+1,\cdots,M+1]}}=0, & \hspace{-40mm}
    \textrm{otherwise}.
  \end{array}
  \right.
\end{equation}
\end{widetext}
According to the results in Eq.~\eqref{eq:eqLambdan} and
\eqref{eq:eqReducedState}, we get the following recursion formula

\begin{eqnarray}\label{eq:eqRecursion}
  & &
    \ket{\psi_{l+1}^{[n,n+1,\cdots,M+1]}}\nonumber\\
  &=&
    \frac{\sqrt{\binom{n-1}{l}}}{\lambda_{l+1}^{[n-1]}}
    \left[\frac{\lambda_{l+1}^{[n]}}{\sqrt{\binom{n}{l}}}
    \ket{0}\ket{\psi_{l+1}^{[n+1,\cdots,M+1]}}\right. \nonumber\\
  & &
    \left.+
    \frac{\lambda_{l+2}^{[n]}}{\sqrt{\binom{n}{l+1}}}
    \ket{1}\ket{\psi_{l+2}^{[n+1,\cdots,M+1]}}\right]
\end{eqnarray}
Comparing Eq.~\eqref{eq:eqRecursion} with the following relation
\begin{eqnarray*}\label{eq:eqRecursionPar}
  & &
    \ket{\psi_{l+1}^{[n,n+1,\cdots,M+1]}}\nonumber \\
  &=&
    \sum_{\alpha_{n},i_{n}}{
    \Gamma_{(l+1)\alpha_{n}}^{[n]i_{n}}\lambda_{\alpha_{n}}^{[n]}
    \ket{i_{n}}\ket{\psi_{\alpha_{n}}^{[n+1,\cdots,M+1]}}},
\end{eqnarray*}
we have
\begin{equation}\label{eq:eqGamma1}
  \Gamma_{(l+1)\alpha_{n}}^{[n]0}=\delta_{(l+1)\alpha_{n}}
  \frac{\sqrt{\binom{n-1}{l}}}
  {\lambda_{l+1}^{[n-1]}\sqrt{\binom{n}{l}}},
\end{equation}
\begin{equation}\label{eq:eqGamma2}
  \Gamma_{(l+1)\alpha_{n}}^{[n]1}=\delta_{(l+2)\alpha_{n}}
  \frac{\sqrt{\binom{n-1}{l}}}
  {\lambda_{l+1}^{[n-1]}\sqrt{\binom{n}{l+1}}}.
\end{equation}
In order to get the MPS form in Eq.~\eqref{eq:eqMPS}, we define that
\begin{equation}\label{eq:eqVn}
  V_{\alpha_{n}\alpha_{n-1}}^{[n]i_{n}}=
  \Gamma_{\alpha_{n-1}\alpha_{n}}^{[n]i_{n}}\lambda_{\alpha_{n}}^{[n]},
  \quad (1<n\leq M).
\end{equation}

After performing $M$ sequential Schmidt decompositions, the states
on the rhs in Eq.~\eqref{eq:eqSDn} can be written as
$\ket{\psi_{\frac{M+1}{2}}^{[M+1]}}=-\ket{0}$ and
$\ket{\psi_{M+1}^{[M+1]}}=\ket{1}$. Checking the above-defined $V$,
we find that the isometry condition
$\sum_{i_{n}}{[V^{[n]i_{n}}]^{\dagger}V^{[n]i_{n}}}=1$ is satisfied.

Until now, we have found that the output state of the general
quantum NOT gate can be expressed as a MPS as in
form~\eqref{eq:eqMPS}. So the sequential quantum NOT gate is
obtainable.

\section{Concluding Remark}

In summary, by applying quantum orthogonalizing transformations
for the state uniformly distributed between two latitudes on the
Bloch sphere, we present a general $1$-to-$M$ quantum NOT gate. The
usual U-NOT gate is a special case and we find out that the optimal
fidelity of the U-NOT gate is consistent with the one studied
in~\cite{BHW-UNG1999}. For another special case, we point out the
relation between the phase-covariant $1$-to-$M$ NOT gate and the
phase-covariant QCM. In the general situation, there is no
``measurement + repreparation" scheme as the U-NOT gate can be
realized via it. Consequently, the generation of entanglement is
unavoidable and the controlled generation of entangled states
becomes very important. To this end, we analyze the NOT gate within a
sequential generation scheme and show that the sequential quantum
NOT gate is feasible.

{\bf Acknowledgement:} This work was supported in part by the
National Basic Research Program of China grant No. 2007CB907900 and
2007CB807901, NSFC grant No. 60725416 and China Hi-Tech program
grant No. 2006AA01Z420.  KLC acknowledges support by the National
Research Foundation and Ministry of Education, Singapore


\end{document}